\documentstyle[aps,epsf]{revtex}
\begin{document}
\large
\title {The Accuracy of Semiclassical Quantization for Integrable Systems}
\author {Saar Rahav\footnotemark[2], Oded Agam\footnotemark[3] and Shmuel Fishman\footnotemark[2] \\ \dag \hskip.1in Department of Physics, Technion, Haifa 32000, Israel \\ \ddag \hskip.1in The Racah Institute of Physics, The Hebrew University, Jerusalem 91904, Israel}
\date{24 May 1999}
\maketitle

\begin{abstract}
The eigenvalues of the Hyperspherical billiard are calculated in the
semiclassical approximation. The eigenvalues where this approximation
fails are identified and found to be related to caustics that approach
the wall of the billiard. The fraction of energy levels for which the
semiclassical error is larger than some given value is calculated 
analytically (and tested numerically) and found to be independent of 
energy. The implications for other systems, in particular integrable ones,
are discussed. 
\end{abstract}
\section {Introduction}

Exact solutions to physical problems are rare. In most cases one has to resort to approximate solutions. In quantum mechanics the Wentzel-Kramers-Brillouin (WKB) method, along with perturbation theory, is probably the most common method used to obtain approximate solutions, (for reviews see~\cite{mount72,heading62,froman65}).
The WKB method is a formal $\hbar$ expansion for the wave function, that expresses its rapid oscillations in the semiclassical limit. 
Using this expansion combined with an appropriate boundary condition results in an
approximate quantization condition.  
Therefore, high order approximations for the wave function can be used to improve the accuracy of the eigenvalues. A systematic way to obtain approximate quantum eigenvalues using WKB was developed first by Dunham~\cite{Dunham32}, was improved by Bender and coworkers~\cite{bender77} and is summarized in~\cite{bender}.
In some cases the resulting series for the eigenvalues converge to the exact 
ones, but generally the resulting expansion will be an asymptotic series. 

A surprisingly small number of papers were devoted to the accuracy of semiclassical
methods, and even less for high orders or high dimensional systems. 
A naive estimate for the accuracy of semiclassical methods is that 
the leading semiclassical approximation is accurate to order $\hbar$ and therefore the
resulting error is of order $\hbar^2$. For example, substitution of the
Van Vleck propagator into Schr\"odingers equation will not solve the
equation exactly, and a reminder of order $\hbar^2$ will result~\cite{gutz89}.  The mean level spacing scales like 
$\hbar^D$, where $D$ is the dimension of the system, thus the relative error scales like $\hbar^{D-2}$ and the semiclassical
method fails for high dimensional systems. 
The prefactor of the semiclassical error is important (in particular for $D=2$), since individual energies
can be found semiclassically only if the error is less than the mean level spacing.
For integrable systems the WKB expansion enables one to find analytic estimates for
the error in the energies and this will be demonstrated for a class
of such systems in this paper. In addition it may shed light also on the accuracy of the semiclassical
approximation for other dynamical systems. The common argument for the failure of 
the semiclassical approximation for chaotic systems in $D>2$ and possibly for $D=2$ was recently challenged by 
Primack and Smilansky~\cite{primack97?}. 
For chaotic systems energy is the only quantum number and this is the only 
identity of a level. The mean level spacing is determined by the smooth Weyl
term. This term can be determined for billiards to a high order in Plancks
constant $\hbar$. Primack and Smilansky introduced a method to evaluate the 
error in the energy levels resulting from the fact that oscillatory terms are
known only to the leading order in $\hbar$. They concluded that the error
in the evaluation of single levels compared to the mean level spacing
diverges at most as $| \ln \hbar |$ in the limit $ \hbar \rightarrow 0$.
Dahlqvist~\cite{dalqvist99} estimated effects of diffraction for the $D=2$ 
Sinai billiard and concluded that ignoring diffraction, as is done
in the standard semiclassical approximation, may result in errors at least of the
order of the mean level spacing.

 Here we will study the much simpler case of
integrable systems when an expression for the energy levels can be obtained
to arbitrary order in $\hbar$ and the levels that are poorly approximated
in the leading order of the semiclassical approximation can be 
characterized.   
An estimate for the error of the semiclassical 
quantization for the Hyperspherical billiard, that is a generalization of 
 the circle billiard, will be obtained. This billiard  has been
subjected to several works regarding the semiclassical accuracy.
Prosen and Robnik explored the error for the energies of the 
circle billiard numerically~\cite{prosen93}. They found that the mean error
increases with the energy, and concluded that the semiclassical approximation
fails in this simple system. Boasman used the boundary integral method to find the
quantization error for several billiards, including the circle billiard
~\cite{boasman94}. The error was found to be a small fraction of the mean level 
spacing for general two dimensional billiards. For the circle billiard he found that for most
of the energies the semiclassical error is a constant, which is a small fraction
of the mean level spacing. There were large errors for some eigenenergies, 
the corresponding eigenfunctions were found to be affected by caustics,
and it was stated that the fraction of these poorly approximated states
decreases with energy.

Following earlier work of Agam~\cite{agam95} we use a certain WKB expansion,
 keeping the classical quantities fixed, 
that  gives identical 
quantization condition to the one obtained from the Debye expansion of the Bessel function.
This expansion enables us to estimate the fraction of 
states where the error in the leading semiclassical approximation exceeds some value.
 The relation to caustics turns out transparent.
For integrable systems this is a better measure then the 
mean error since some eigenenergies,  that can be clearly characterized,  may have extremely large errors.
The reason is that for integrable systems, contrary to chaotic ones,
there are other quantum numbers in addition to the energy.
These help to characterize the groups of quantum eigenstates that are badly
approximated in the leading order of the semiclassical approximation.

The semiclassical expansion for the eigenvalues of the Hyperspherical billiard
to second order in $\hbar$ is derived in Sec.~\ref{section2}, with the help
of the standard semiclassical expansion, as well as the Debye expansion of
the Bessel functions. In Sec.~\ref{section3} the density of eigenvalues
where the semiclassical approximation is poor (a term that is defined in that
section) is derived analytically and tested numerically. The results are
summarized in Sec.~\ref{section4} and the implications for other systems
are discussed.

\section{The WKB expansion for the Hyperspherical billiard}
\label{section2}
\setcounter{equation}{0}

In this section the semiclassical quantization for the Hyperspherical billiard will
be developed up to second order in $\hbar$. First a short introduction to a systematic
WKB expansion for one dimensional problems will be given. Then the Debye expansion for the Bessel function will be used to derive the second order quantization condition. This derivation is much simpler than the derivation that uses a systematic WKB expansion.
In order to verify that the quantization condition derived from the Debye expansion of the Bessel function is indeed a result of a semiclassical
expansion a WKB series will be developed for the radial wave function. Quantization of this
series indeed gives the same quantization condition that was obtained using the
Debye series (at least up to second order). 

 A scheme for quantization using high order WKB expansion is known, at least for an ordinary differential equation such as the equations found for
separable Hamiltonians~\cite{Dunham32,bender77,bender,robnik97}. For one dimensional systems the Schr\"odinger equation is 
\begin{equation}
 \left[ - \frac{\hbar^2}{2 m} \frac{d^2}{dx^2} + V(x) \right] \psi (x) = E \psi (x) .
\label{oned} 
\end{equation}
The wave function is written in the form
\begin{equation}
 \psi (x) = A \;  \exp \left( \frac{i}{\hbar} \sigma (x) \right) ,
\end{equation}
and the phase is expanded in powers of $\hbar$
\begin{equation}
\sigma (x) = \sum_{k=0}^{\infty} \left( \frac{\hbar}{i} \right)^k  \sigma_{k}
 (x) 
\label{expansion}
\end{equation}
while $A$ is a normalization constant.
The Schr\"odinger equation can be solved order by order in $\hbar$ leading
to a recursion relation for the expansion functions~\cite{Dunham32,bender77,bender}
\begin{eqnarray}
{\sigma'}_{0}^{2} = 2m (E - V(x)) \nonumber \\ \sum_{k=0}^{n} \sigma'_{k} 
\sigma'_{n-k} + \sigma''_{n-1} = 0.
\label{recursion} 
\end{eqnarray}
The quantization condition is derived by demanding that the wave function is 
single valued
\begin{equation}
\label{singleval}
\sum_{k=0}^{\infty} \left( \frac{\hbar}{i} \right)^k  \oint d \sigma_k = 2 \pi \hbar n .
\end{equation}
The first odd term $\sigma_1 $ has the form of a logarithmic derivative,
$ d \sigma_1 = - \frac{1}{4} d (\ln \sigma_0^{'2} ) $. Each turning point
gives a simple zero of $\sigma_0^{'2} $ , and the
contour integral for $\sigma_1$ counts these zeros and results in the Maslov index. All of the other odd
terms are real, and usually without cuts in the complex $x$ plane. 
Thus these
terms will not contribute to the quantization condition~\cite{bender77}. 
Therefore, the quantization condition becomes
\begin{equation}
\label{quantcond1}
\sum_{k=0}^{\infty} \left( \frac{\hbar}{i} \right)^{2k} \oint d \sigma_{2k} = 2 \pi \hbar(n+m/4)
\end{equation}
where m is the Maslov index to be evaluated later (for the simple problem with two
turning points, $m$ is just $2$, the number of zeros of $\sigma_0^{'2} $)   . The fact
that only even terms contribute to the quantization makes this quantization
method efficient since the correction to the eigenvalues will be smaller by $\hbar^2$
instead of $\hbar$. This will motivate us transform the radial differential 
equation
to a form without a first derivative.

The specific problem that will be studied in this work is the D dimensional 
Hyperspherical billiard, namely a free particle inside a D dimensional
ball ($ R^{2} > \sum_{i=1}^{D}  x_{i}^{2}$) with Dirichlet boundary conditions.
Generalized spherical coordinates will be most convenient to solve the Schr\"odinger equation.
The D dimensional Hamiltonian reduces to the Laplacian operator 
\begin{equation}
{\cal H} = - \frac{\hbar^2}{2 m}\left( \frac{\partial^{2}}{\partial r^{2}}+\frac{D-1}{r} \frac{\partial}{\partial r}+\frac{\Delta_{s}}{r^{2}} \right) 
\end{equation}
with the boundary condition $\psi(R)=0$ on the wave functions.
The generalized angular momentum operator is  $\Delta_{s}$ and its
eigenvalues are $-l(l+D-2) $ \cite{avery89}.
The radial and angular variables can be separated as $\Psi= R(r) \chi(\Omega)$.
For the Hyperspherical billiard the radial Schr\"odinger
equation is
\begin{equation}
	{\cal R}'' (r) + \frac{D-1}{r} {\cal R}' (r)
	 - \frac{l(l+D-2)}{r^2} {\cal R} (r) + 
 	\frac{2 m E}{\hbar^2} {\cal R} (r) =0
\label{radialeq}
\end{equation}

The exact solution can be easily obtained using the
fact that the radial equation~(\ref{radialeq}) can be reduced to the Bessel
equation of order $\nu$, namely
\begin{equation}
u^2 {\cal J}'' + u{\cal J}' + (u^2 - \nu^2) {\cal J} = 0
\end{equation}
where $R(u)= u^{\frac{2-D}{2}}{\cal J}(u)$, $\nu= l + \frac{D-2}{2} $ 
and $u= \sqrt{ \frac{2 m E}{\hbar^2} } r$. The 
solutions are $ {\cal J}_{\nu}$, the Bessel functions of the first kind of order $\nu$, and the quantization condition is:
\begin{equation}
{\cal J}_{\nu} \left( \sqrt{ \frac{2 m E}{\hbar^2}} R \right) =0
\label{exactquan} 
\end{equation} 
This result is exact.
The dimensionless variables 
\begin{eqnarray}
 z & = & \frac{ \sqrt{2 m E} R}{L_{sc}} \nonumber \\ \nu & = & \frac{L_{sc}}{\hbar} =l+ \frac{D-2}{2} 
\label{nodimen}
\end{eqnarray} 
will be introduced for simplicity.
 It is obvious that $z>1$.

 In order to make contact with the semiclassical expansion it is useful
to introduce a large order expansion of the Bessel function since
for small $\hbar$, the order $\nu$ is large. This is the Debye
asymptotic expansion.

 \begin{eqnarray}
	{{\cal J}_{\nu} (\nu z)}_{\nu \rightarrow \infty}
        \sim \left(\frac{2}{\pi\nu \sqrt{z^{2}-1}}\right)^{\frac{1}{2}}
	\left[ \cos \zeta \sum_{m=0}^{\infty} \frac{(-1)^{m} \Gamma (2m+\frac{1}
         {2}) a_{2m}}{\Gamma (\frac{1}{2})}
	\left(\frac{2}{\nu \sqrt{z^{2}-1}}\right)^{2m} \right.+\nonumber \\  
	\left.+ \sin \zeta \sum_{m=0}^{\infty} \frac{(-1)^{m} \Gamma (2m+\frac{3}
         {2}) a_{2m+1}}{\Gamma (\frac{1}{2})}
	\left(\frac{2}{\nu \sqrt{z^{2}-1}}\right)^{2m+1} 
	\right]
\label{debyexp}
 \end{eqnarray}
Where $ \zeta=\nu\left[\sqrt{z^{2}-1}-\arccos \left(\frac{1}{z}\right)\right]-\frac{\pi}{4}$ and  
 $a_{0}=1$  , while
  $a_{1}=\frac{1}{8}+\frac{5}{24} \frac{1}{z^{2}-1}$, and
 $a_{2}=\frac{3}{128}+\frac{77}{576}\frac{1}{z^{2}-1}+\frac{385}{3456} \frac{1}{(z^{2}-1)^{2}} \;$,
 $ \cdots$
\cite {watson58}.

Semiclassical quantization, using the Debye expansion,
is performed in orders of $\hbar$, by requiring that the wave function will have a zero at $r=R$,
order by order in $\frac{1}{\nu}$. Thus to the 
second order the quantization condition is:
\begin{equation}
\cos \zeta + \frac{1}{\nu \sqrt{z^2 -1 }} \left( \frac{1}{8} + \frac{5}{24}
\frac{1}{z^2-1} \right) \sin \zeta = 0 . 
\label{debye2}
\end{equation}

First we solve for the first order condition and correct it perturbatively.
 Define $\zeta_0 =\zeta (z_0)$ as the leading
order solution. It satisfies
\begin{equation}
\label{eqcos}
\cos \zeta_0 = 0 . 
\end{equation}
The second order solution can be expanded around the first order
one: $ z = z_0 + \delta z $. The term $\delta \zeta = \zeta (z_0 + \delta z)- \zeta ( z_0 )$ is of
the order $\frac{1}{\nu}$ and therefore small. Substituting in~(\ref{debye2}) one finds
\begin{equation}
\label{eqcos2}
\delta \zeta = \frac{1}{\nu \sqrt{z^2-1}} \left( \frac{1}{8} + \frac{5}{24} 
\frac{1}{z^2-1} \right) .   
\end{equation}
This last expression can be given in terms of the first order solutions $z_0$
or the second order ones since the difference is of higher order
in $\frac{1}{\nu}$. The last step is to rewrite the first order condition
as $\cos \zeta_0 = \cos ( \zeta -\delta \zeta ) = 0$. This results in 
\begin{equation}
\label{sem2a}
\nu \left( \sqrt{z^2 - 1} -\arccos \left( \frac{1}{z} \right) \right) -
\frac{1}{\nu \sqrt{z^2 -1 }} \left( \frac{1}{8} + \frac{5}{24} \frac{1}{z^2 -1}
\right) = \pi \left( n + \frac{3}{4} \right) . 
\end{equation}   
This is the second order semiclassical quantization condition. For every pair
of quantum numbers $(n,\nu)$ one can solve for $z$ and obtain $E_{sc}(n,\nu)$.
Since the quantization condition is determined by only two 
quantum numbers, the problem
behaves like a system of two 
degrees of freedom , if one chooses to ignore degeneracies.
The reason for denoting this quantization condition as a second order one will become clear 
shortly.

We turn now to apply the systematic semiclassical expansion to~(\ref{radialeq}).
Our goal is to show that the quantization condition that was obtained 
from asymptotic expansion of Bessel functions
coincides with the one obtained from a WKB expansion.
This equation has a first order derivative and therefore does not have the form
of the one dimensional Schr\"odinger equation~(\ref{oned}). To exploit the 
formal WKB expansion described above one has to eliminate
the first order derivative from the equation. Substitution of
\begin{equation}
{\cal R} (r) = r^{\frac{1-D}{2}} \phi (r) 
\end{equation}  
in~(\ref{radialeq})  leads to the following equation for $\phi (r)$
\begin{equation}
 - \frac{\hbar^2}{2 m} \phi'' (r) + \frac{\hbar^2}{2 m r^2} \left[ ( l + 
 \frac{D-2}{2} )^2 - \frac{1}{4} \right] \phi (r) = E \phi (r) . 
\label{nofirsteq}
\end{equation}

Thus after this substitution the resulting equation will be
similar to that of a system with one degree of freedom and the WKB method
will give better results. 
The angular momentum that appears in this equation is the semiclassical one 
(see~(\ref{nodimen}) and App.~\ref{appa}),
as opposed to the exact angular momentum $\hbar \sqrt{l (l+D-2)}$ as will be explained in what follows.
 These differ by an extra term that vanishes in
the limit $\hbar \rightarrow 0$. That gives one some freedom in defining the semiclassical variables. Thus, for example, we can define the limit as $\hbar \rightarrow 0$ while $E$
and $ L^2=\hbar^2 l (l+D-2)$ are constants. This prescription for
the semiclassical limit is problematic, 
 since the form of the resulting leading order 
wave function near zero and at infinity differs than from the one of the exact
wave function. This problem is known for the spherical case ($D=3$), where the
exact angular momentum is replaced by the semiclassical one, a modification
introduced by Langer~\cite{langer37,mount72}. For the $D$ dimensional generalized angular 
momentum the analog is the replacement of $(l+ \frac{D-2}{2})^2 - \frac{1}{4}$
by $(l+ \frac{D-2}{2})^2$. 
 Therefore, we define the semiclassical limit
as the limit where $ \hbar \rightarrow 0$ while $E$ and $L_{sc}=\hbar  (l+\frac{D-2}{2})$ 
are constant.
The difference between these definitions of the angular momentum
vanishes in the limit $ \hbar \rightarrow 0$.  As a result, we will apply the WKB
approximation to the 
following equation
\begin{equation}
 - \hbar^2 \phi'' (r) = 2 m \left( E - \frac{L_{sc}^{2} - \frac{\hbar^2}{4}}{2 m r^2} \right) 
\phi (r) .
\end{equation}
Using the expansion~(\ref{expansion}) for the wave function we find that the recursion 
relation~(\ref{recursion}) changes to
\begin{eqnarray}
 {\sigma'}_{0}^{2} (r) = 2 m \left( E - \frac{L_{sc}^2}{2 m r^2} \right) \nonumber \\
  \sum_{k=0}^{n} \sigma'_{k} 
\sigma'_{n-k} + \sigma''_{n-1} = - \frac{1}{4 r^2} \delta_{n,2} 
\label{newrecursion} 
\end{eqnarray}

The first order contribution to the quantization is obtained from the $0'th$ order
 of~(\ref{newrecursion})
\begin{equation}
 \oint \sigma'_{0} dr = 2 \int^{R}_{\sqrt{ \frac{L_{sc}^2}{2 m E}}} 
 \frac{1}{r} \sqrt{2 m E r^2 - L_{sc}^2} dr = 2 \hbar \nu \left[ \sqrt{z^2 -1} - \arccos \left(
\frac{1}{z} \right) \right] . 
\label{firstorderint}
\end{equation}
The variable z (see Eq.~(\ref{nodimen}))
is the ratio between the radius of the billiard and the minimal distance
of the classical orbit from the origin, or the radius of the caustic.
Therefore, if z is close to $1$
then the state is more affected by caustics and the semiclassical 
approximation deteriorates. The reason is that the wave function of these 
states is close both
to the caustic and to the hard wall where the pure semiclassical approximation fails.
 The next term will give us the 
Maslov index. The integration over 
$\sigma'_1$ will give an extra phase of $ \frac{\pi}{2}$ due to the turning
point at $r=\sqrt{ \frac{L_{sc}^2}{2 m E}}$ (resulting in a zero
of $\sigma_0^{'2}$ as discussed after~(\ref{singleval})). We have to include also the hard wall at $r=R$ 
. To do this correctly one has to build the wavefunction for the sum of
to solutions in such way that they cancel each other at
the wall. The result is that the reflected wave acquires an 
extra phase of $\pi$, therefore the total Maslov index is just 3.
Consequently the quantization condition in the leading order is 
 \begin{eqnarray}
	\nu\left[\sqrt{z^{2}-1}-\arccos \left(\frac{1}{z}
        \right)\right] =
	\pi\left(n+\frac{3}{4}\right).
 \label{sem}
 \end{eqnarray} 
In App.~\ref{appa} it is derived in the framework of the ``old quantum theory''
by direct action quantization.

 Our goal is now to obtain the second order semiclassical
condition. A straight forward calculation leads to
\begin{equation}
\label{sigma2}
 \sigma'_{2} (r) = \frac{ L_{sc}^2 ( L_{sc}^2 - 12 m E r^2)}{8 r ( 2 m E r^2 - L_{sc}^2 )^{5/2}} - \frac{1}{8 r \sqrt{2 m E r^2  - L_{sc}^2} } . 
\end{equation}
The contour integral over $ \sigma'_2 $ is calculated in App.~\ref{appb}, and the result is:
\begin{equation}
\label{oint2}
\oint \sigma'_2 dr = \frac{1}{4 \hbar \nu \sqrt{z^2 -1}} + \frac{5}{ 
12 \hbar \nu \left( z^2 -1 \right)^{3/2}} .
\end{equation} 
Using the quantization condition~(\ref{quantcond1}) up to the second order
\begin{equation}
\oint \left( \sigma'_0 - \hbar^2 \sigma'_2
 \right) dr = 2 \pi \hbar \left(n + \frac{m}{4}\right) 
\end{equation}
and~(\ref{firstorderint}) one obtains the second order quantization condition
\begin{equation}
\label{sem2}
\nu \left( \sqrt{z^2 - 1} -\arccos \left( \frac{1}{z} \right) \right) -
\frac{1}{\nu \sqrt{z^2 -1 }} \left( \frac{1}{8} + \frac{5}{24} \frac{1}{z^2 -1}
\right) = \pi \left( n + \frac{3}{4} \right) . 
\end{equation}   
It is identical to (\ref{sem2a}), obtained from the Debye expansion.
If $L_{sc}$ is kept fixed in the semiclassical limit then the limit $\hbar \rightarrow 0$ is
equivalent to $\nu \rightarrow \infty$. Note that if $z$ is not too close to $1$ the first term is much larger then the second one and dominates the result.

It was demonstrated explicitly that the Debye series and the WKB method lead to the same quantization condition up to the second order. Furthermore, these expansions are valid in
the {\em same} limit. Therefore we expect that these expansions give
identical results in any order. This enables one to use the Debye 
asymptotic expansion, in this case, to obtain estimates for the semiclassical error.
 This
will be the subject of  the next section.

Before we continue, it is worthwhile discuss briefly the semiclassical
angular momentum $L_{sc}$. In the App.~\ref{appa}
we calculated the semiclassical eigenvalues of the generalized angular momentum using 
first order WKB.
One may ask whether it is possible to obtain the exact form of the angular momentum eigenvalues
from a WKB expansion. It turns out that this is indeed the case. The WKB approximation for
the 3 dimensional angular momentum was treated by Robnik and Salasnich~\cite{robnik97b}.
They computed the first several terms in the series for the angular momentum
eigenvalues, and conjectured the general form of this
series. This conjecture was later proved by Salasnich and Sattin~\cite{salasnich97}.
 The generalization for high dimensional systems is not difficult and requires only minor changes in the differential equation and the WKB series. The details are presented in App.~\ref{appc}.
The convergent series leads to the correct eigenvalues for the generalized angular
momentum (namely $ \hbar \sqrt{l (l+D-2)} $), while the first order 
in the WKB expansion is $L_{sc}=\hbar (l+\frac{D-2}{2})$. 

\section{The semiclassical error for Hyperspherical billiards}
\label{section3}

The leading semiclassical eigenvalues correspond to the zeros of
 ${\cal J}^{(o)}_{\nu}(\nu z)$, the leading
term  in the Debye asymptotic expansion. 
 Here ${\cal J}^{(i)}_{\nu}(\nu z)$ denotes the i'th  order term in this expansion.
Let $\Delta z =z(E_{ex})-z(E_{sc})$, then to the leading order in
this asymptotic expansion 
 $\Delta z = \frac{{\cal J}_{\nu}^{(0)}(\nu z)-{\cal J}_{\nu}(\nu z)}{{\cal J}_{\nu}^{'(o)}(\nu z)}\simeq-\frac{{\cal J}_{\nu}^{(1)}(\nu z)}{{\cal J}_{\nu}^{'(o)}(\nu z)} $
\cite {boasman94,agam95}, where ${\cal J}_{\nu}^{'(o)}(\nu z) $ is the first 
derivative of  ${\cal J}_{\nu}^{(o)}(\nu z) $ with respect to $z$ and from
~(\ref{debyexp}) one finds, $\Delta z = \frac{\delta \zeta}{\left( \frac{d \zeta_0}{d z_0} \right)}$ where $z_0$ is the leading order solution satisfying ${\cal J}^{(0)}_{\nu} (\nu z_0) = 0$
(see~(\ref{eqcos}) and~(\ref{eqcos2}). The result can be
obtained also directly from these equations).

If the difference $\Delta z$ is small then 
 $\frac{\Delta z}{z}=\frac{\Delta E}{2E}$.
 Using~(\ref{nodimen}),
 \begin{eqnarray}
	|\Delta E| \simeq \frac{\hbar^{2}}{mR^{2}} \frac{z^{2}}{z^{2}-1}
	\left[\frac{1}{8}+\frac{5}{24} \frac{1}{z^{2}-1}\right]
 \label{eq:diffe}
 \end{eqnarray}
that is correct in the leading order in $\hbar$.
Note that the error $\Delta E$ is independent of the number of degrees of freedom D.
In the limit $ z \rightarrow 1$ the expression diverges. This divergence is related to fact that in this limit the caustic approaches the wall and 
 the classical trajectories on the quantized torus
are always adjacent to the caustic and to the hard wall.
 The approximation is best in the limit 
$z \rightarrow \infty$. In this limit the angular momentum contribution
to the energy is negligible, and the problem is effectively one dimensional.

To estimate the accuracy of the approximation
it may be more meaningful to
measure the error in units of the mean level spacing $\Delta$. In the $D$ dimensional
 Hyperspherical billiard the leading order in the mean level spacing is given by
the Weyl formula
 \begin{eqnarray}
	\Delta \simeq \frac{(2\pi)^{D} \hbar^{D}}
	{mR^{2} V^{2}_{D} D L_{sc}^{D-2}}
	\frac{1}{z^{D-2}}
 \label{eq:delta1}
 \end{eqnarray}
where $V_{D} = \frac{\pi^{\frac{D}{2}}}{\Gamma(1+\frac{D}{2})}$
is the volume of the $D$ dimensional sphere of unit radius.. With (\ref{eq:diffe}) and~(\ref{eq:delta1})
the semiclassical error is: 
 \begin{eqnarray}
	\frac{|\Delta E|}{\Delta} \simeq \frac{D L_{sc}^{D-2}}
	{\Gamma^{2}(1+\frac{D}{2}) 2^{D}} \frac{z^{D}} {z^{2}-1}
	\left[ \frac{1}{8}+\frac{5}{24} \frac{1}{z^{2}-1} \right]
        \frac{1}{\hbar^{D-2}}
 \label{error1}
 \end{eqnarray}
In the calculation of the mean level spacing $\Delta$, the levels were weighted with
their degeneracy, leading to a large level density, resulting in a very small mean
level spacing. The divergence of $\frac{ | \Delta E |}{\Delta}$ for $D>2$ in the limit 
$\hbar \rightarrow 0$ ( $L_{sc}$ and $z$ fixed) and
$ z \rightarrow \infty $ ( $L_{sc}$ and $\hbar$ fixed) is a result of the rapid increase of the number of levels
in these limits. 

The Hyperspherical billiard exhibits a high degree of symmetry. Therefore
 there are only two quantum numbers, $(n,l)$, that determine the quantization
condition. As a result, if levels are not weighted by their degeneracies
 the density of the degenerate energy 
levels is quasi two dimensional. Consequently, if the number of states that 
are affected by caustics is small than one can use WKB to determine single
energy levels. Knowledge of
the relative number of states that are badly approximated is crucial here. If 
such states are common than the semiclassical approximation will fail. 
In what follows the fraction of states that are badly approximated by WKB will be estimated.

One can decompose the total spectrum using the angular momentum quantum number. Each 
sub-spectrum is a series of eigenvalues that depend on the quantum number $n$.
The density of these states with respect to the energy can be approximated from
the quantization condition~(\ref{sem})

\begin{equation}
\rho_n \simeq \left( \frac{ \partial n}{\partial E } \right) = 
\frac{\nu}{2 \pi E} \sqrt{ \frac{E - \epsilon(\nu)}{\epsilon(\nu)}} 
\end{equation}
where $\epsilon(\nu)= \frac{\hbar^2 \nu^2}{2 m R^2}$. The decomposed spectrum is not 
very interesting, since our purpose is to find what is the total fraction of states that are baldy approximated irrespective of
the other quantum numbers.  
For this, the spectrum can be constructed by adding the contributions from different angular momenta. It is
important to understand that when we sum over $\nu$ the energy is kept fixed, thus we
actually sum over the different quantum states $(\nu , n)$ with energy near $E$.
If the
sum is over all the possible values of $\nu$ then one is counting all the possible states
and will get the total density of states. If one wants to sum only the states that are
badly approximated one should sum only the states that satisfy the condition $1 \leq z \leq
1/\alpha$, where $\alpha$ is a measure for the maximally allowed error.
From~(\ref{eq:diffe}) it is clear that $|\Delta E|$ is monotonically decreasing with $z$.  
To find the density of badly approximated states one sums only values of $\nu$ satisfying this inequality.
Replacing the restricted sum $\sum^{'}$ by an integral and changing variables
to $\epsilon$, leads to 

\begin{equation}
\rho \simeq {\textstyle \sum_{\nu}^{'}} 
\frac{\nu}{2 \pi E} \sqrt{ \frac{E - \epsilon(\nu)}{\epsilon(\nu)}}
\simeq \frac{m R^2}{2 \pi \hbar^2} \int_{\alpha^2 E}^E \sqrt{\frac{ E - \epsilon}{\epsilon}}
\frac{d \epsilon}{E} \; . 
\end{equation} 
The resulting density of badly approximated states is
\begin{equation}
 \rho \simeq \frac{m R^2}{2 \pi \hbar^2} \left[ \arccos \alpha - \alpha \sqrt{1- \alpha^2} 
\right] .
\label{baden}
\end{equation}
The result depends only on $\alpha$, which is a measure of the error but does not depend on energy. The density of badly approximated states does not change with the energy!! Thus, the probability
that a state is badly approximated by the WKB method does not depend upon 
the energy of the state. It is important to notice that this expression is useful
only in the limit $\alpha << 1$. When  $\alpha \simeq 1$ the connection between the allowed
error and $\alpha$ is not accurate since the various terms in the Debye series are of comparable
magnitude and the expansion cannot be terminated. Indeed, when all of the states are included, (for $\alpha = 0$ any state is considered to be badly approximated),
the density of all the states is $\frac{m R^2}{4 \hbar^2}$,
as expected since this is the total density of states according
to the Weyl formula in two dimensions (the double degeneracy of states with positive and negative angular momentum quantum numbers
is ignored here). 

\begin{figure}[b]
\epsfxsize 4.in
\epsfysize 3.5in
\epsfbox[-100 30 350 500 ]{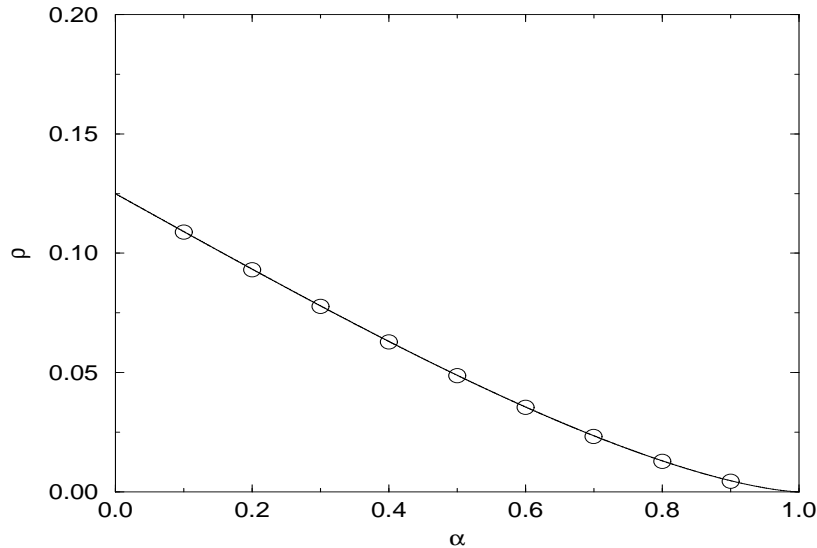}

\caption{The density of badly approximated states. (solid line - the approximation, circles - numerical results)  \label{fig:figu1}} 	
\end{figure}

To verify these results we computed the first 125336 exact and semiclassical levels in the D=4 dimensional Hyperspherical
billiard numerically. A convenient set of units is $ 2 m = \hbar = R = 1$. 
The error in each eigenvalue was computed directly from comparison between the exact energy levels computed from~(\ref{exactquan}) and  their semiclassical 
approximation computed from~(\ref{sem})
. A set of $\alpha$ values was chosen. Each value of $\alpha$ defines a semiclassical
error by~(\ref{eq:diffe}), with $z=\frac{1}{\alpha}$. Then, the density of states with error larger
than this semiclassical error was computed by counting the number of such states
in an energy interval which is much larger than the mean level spacing but small enough not to smear any possible energy dependence. This number is just
the density of badly approximated states times the width of
the energy interval. 
For each $\alpha$ the density of badly approximated states was found to be constant when the energy
was changed, as expected from~(\ref{baden}). The value of this density was compared
to the density predicted by~(\ref{baden}),   
and is presented in~Fig. \ref{fig:figu1}. The approximation
~(\ref{baden}) was found to be excellent.

It was found that using the fact that the system is integrable and
the knowledge of its degeneracy, we can treat the spectrum
of the Hyperspherical billiard as two dimensional.
The density of badly approximated states was calculated and was found to be energy independent contrary to a previous statements~\cite{boasman94}. Only in the limit where the angular momentum is negligible there 
are no large errors in quantization, since the system is effectively one dimensional.

\section{Discussion and conclusions}
\label{section4}

The accuracy of the semiclassical approximation was studied for the
Hyperspherical billiard. The approximation was found to fail for energies
corresponding to states that are localized on such tori, that the caustic
is close to the wall. As the caustic approaches the wall the region in space
where the semiclassical approximation for the wave function is justified
shrinks, and the quality of this approximation deteriorates. The semiclassical
approximation for eigenvalues can be arbitrary bad (for $z \simeq 1$, in our case).
For this reason the average semiclassical error is meaningless since it may
be dominated by few large contributions. It is much more reasonable
to calculate the fraction of badly approximated states, namely the fraction
of states where the semiclassical error exceeds some value. The density of 
such states was found to depend only on one parameter that measures
the ratio between the angular momentum and its maximal possible classical
value for a given energy (the parameter was denoted by $\alpha$, and
the regime where the semiclassical approximation fails is $1 < z < \frac{1}{\alpha}$).
The semiclassical error~(\ref{eq:diffe}) depends only on $z$ and not on the
energy. Therefore we concluded that the density of poorly approximated
eigenvalues characterized by $\alpha$ is independent of energy. The 
semiclassical approximation does {\em not} improve as the energy increases,
contrary to common belief (stated for example in~\cite{boasman94}).
Lazutkin found that convex billiards have tori of the KAM type along the boundary.
He also proved that it is possible to use a WKB like method to obtain approximate
solutions and energy levels on these tori~\cite{lazutkin93}. 
For these states there is a caustic near a hard wall and we expect that the energy 
levels will be badly approximated.

For integrable systems there are symmetries and the situation where 
degeneracies exist is quite typical. Since the symmetries are respected
by the semiclassical approximation also the degeneracies are preserved
and the levels of the semiclassical approximation have the same
degeneracy as the exact ones.
Therefore it is meaningless to compare the semiclassical error to the
mean level spacing that is inversely proportional to the total number of
energy levels. It is much more meaningful to consider the error within
each group characterized by some symmetry and therefore by an additional
quantum number.  Since the existence
of quantum numbers in addition to energy and of caustics are generic
for integrable systems we expect some of the results found in this work
to be generic for such systems. It is expected that the semiclassical
approximation will fail for well defined groups of states and the
fraction of such states, rather than the mean semiclassical error is
a good measure of the quality of the semiclassical approximation.
To our understanding in general the density of badly approximated
states may depend on energy, contrary to the situation in the present work.
The characterization of this dependence should be the subject of further
studies.

The results of this work may be relevant for mixed systems, for quantum states
that are localized in the regular regions. For chaotic systems and for
the chaotic component of mixed systems, there are no caustics and the wave functions
usually spreads over all the chaotic region. The energy is the only identity of an eigenstate. Therefore the results of this work may not be of
direct relevance for such systems and the mechanism for the 
destruction of the semiclassical approximation is different there. Some
corrections to the trace formula for chaotic systems may be of a similar origin.

Alonso and Gaspard computed the correction to the Gutzwiller trace formulae 
for billiards~\cite{alonso93}. The correction is complicated but some of its parts have
geometrical meaning.
One term is proportional to the sum $\sum_{i} \frac{1}{C_i}$ where $C_{i}$ is the chord length.
Another contribution is proportional to $\sum_{i} \frac{1}{R \cos^3 \phi}$ where $R$ is
the radius of curvature of the wall and $\phi$ is the angle between the 
orbit and the normal to the wall at the incident point . When a caustic
is close to a hard wall the typical chord length will be small and also  
$\cos \phi$. Thus the correction to the leading order contribution of periodic orbits
will diverge and the semiclassical quantization may fail.
Therefore we expect that semiclassical methods will give poor results
for contributions
of orbits that are adjacent to a hard wall. 
For chaotic systems, because of ergodicity a typical
orbit will not have a large fraction of chords near the wall of a billiard
therefore this
correction will not be dominate. One can expect that in such systems the semiclassical
error will
fluctuate around some average without extreme deviations.
As is clear from the present work the situation will be different for 
integrable and mixed systems.

\begin{acknowledgments}

It is our pleasure to thank B. Eckhardt, M. Robnik and
U. Smilansky for interesting discussions. We would like to thank V. Kravtsov
for the hospitality during the workshop on Quantum Chaos and Mesoscopics
(Trieste, Aug 98) where part of this work was done. 
This research was supported in part by the U.S.--Israel Binational Science
Foundation (BSF),
by the Minerva Center for Non-linear Physics of Complex Systems, by the
Israel Science Foundation,
by the Niedersachsen Ministry of Science (Germany) and by the Fund for
Promotion of Research at the Technion.
\end{acknowledgments}

\begin{appendix}
\section{Leading order quantization}

\label{appa}
In this Appendix we use the ``old quantum theory'' in order to obtain the semiclassical quantization condition~(\ref{sem}). The method uses the classical momenta 
for quantization of action integrals.
This result is not new but it may help to clarify the
form of the semiclassical limit that was used in the systematic WKB
expansion.  
Our system is a $D$ dimensional spherical billiard and it is convenient
to describe it using a Hyperspherical coordinate system. The transformation between
the Hyperspherical and Cartesian coordinates is:

\begin{eqnarray}
 x_D & = & r \cos \xi_{D-1} \nonumber \\
 x_{D-1} & = & r \sin \xi_{D-1}\cos \xi_{D-2} \nonumber \\
 x_{D-2} & = & r \sin \xi_{D-1}\sin \xi_{D-2} \cos \xi_{D-3} \\
\vdots & & \vdots \nonumber \\
 x_1 & = & r \sin \xi_{D-1}\sin \xi_{D-2} \cdots \sin \xi_2 \sin \xi_1 \nonumber
\end{eqnarray}
It is easy to prove that the Hamiltonian of a free particle in the spherical
coordinates is given by
\begin{eqnarray}
{\cal H}_D & = & \frac{1}{2 m} \left( p_r^2 
+ \frac{p^2_{\xi_{D-1}}}{r^2} +\frac{p^2_{\xi_{D-2}}}{r^2 \sin^2 \xi_{D-1}} +
\frac{p^2_{\xi_{D-3}}}{r^2 \sin^2 \xi_{D-1}\sin^2 \xi_{D-2}} + \cdots 
+\frac{p^2_{\xi_{1}}}{r^2 \prod_{i=2}^{D-1} \sin^2 \xi_i} \right)
\end{eqnarray}
This Hamiltonian satisfies the Staeckel conditions~\cite{goldstein}, therefore
Hamilton's characteristic function is completely separable
\begin{equation}
 W(q) = \sum_i W_i(q_i) 
\end{equation}
The Hamilton-Jacobi equation in these coordinates is
\begin{eqnarray}
 \left( \frac{\partial W_r}{\partial r} \right)^2 +
 \frac{1}{r^2} \left[ \left( \frac{\partial W_{\xi_{D-1}}}{\partial \xi_{D-1}}
 \right)^2 + \frac{1}{\sin^2 \xi_{D-1}} \left[
 \left( \frac{\partial W_{\xi_{D-2}}}{\partial \xi_{D-2}}\right)^2 \right. \right.
 & & \nonumber \\ \left. \left.
 + \frac{1}{ \sin^2 \xi_{D-2}} \left[ \cdots \left[
 \left( \frac{\partial W_{\xi_{2}}}{\partial \xi_{2}} \right)^2 +
 \frac{1}{ \sin^2 \xi_{2}} 
 \left( \frac{\partial W_{\xi_{1}}}{\partial \xi_{1}} \right)^2
	\right] \cdots \right] \right] \right] & = & 2 m E .
\end{eqnarray}
The brackets in this equation must be constants of motion since they depend upon different
variables. Since the constants are positive we will define $\alpha^2_i$ as the 
value of the $i$th brackets.
  We can write the Hamiltonian as
\begin{equation}
{\cal H}= \frac{p^2_r}{2 m} + \frac{L^2}{2 m r^2}
\end{equation}
where $L=\alpha_{D-1}$ is the generalized angular momentum. 
We will calculate $\alpha_{D-1}$ by calculating the actions related to the
angular variables. The first coordinate is cyclic thus
\begin{equation} 
{\cal S}_1 = 2 \pi \alpha_1 .
\end{equation}
All the other actions are similar and are computed in~\cite{goldstein}.
\begin{equation}
{\cal S}_i = \oint \frac{\partial W_{\xi_i}}{\partial \xi_i} d \xi_i =
 \oint \sqrt{\alpha_i^2 - \frac{\alpha^2_{i-1}}{\sin^2 \xi_i}}  d \xi_i =
 2 \pi (\alpha_i-\alpha_{i-1}) .
\end{equation}
The WKB quantization is done by quantization of the actions 
 ${\cal S}_i = 2 \pi \hbar (l_i + \gamma_i/4)$ where $\gamma_i$ is the Maslov index
 of the action. Now the angular momentum is given by 
\begin{equation}
L = \alpha_{D-1} = \frac{1}{2 \pi} \sum_i {\cal S}_i = \hbar 
\left( \sum_{i=2}^{D-1} (l_i + \frac{1}{2}) + l_1 \right) =
\hbar \left( l + \frac{D-2}{2} \right) .
\label{semiangmom}
\end{equation}
This is just the semiclassical angular momentum $L_{sc}$.

One additional quantization is needed in order to obtain the energy levels
\begin{equation}
{\cal S}_r= \oint p_r dr = 2 \int^{R}_{\sqrt{\frac{L^2}{2 m E}}} \sqrt{ 2 m E -
\frac{L^2}{r^2}} d r
\end{equation}
where $R$ is the radius of the Hyperspherical billiard. Here the Maslov index is $3$, since the contribution from the turning point is $1$ and
the hard wall contributes $2$. A 
straight forward calculation leads to the
quantization condition:
\begin{equation}
{\cal S}_r= 2 \left[ \sqrt{2 m E R^2 - L^2} -L \arccos \sqrt{ \frac{L^2}{2 m E R^2}}
\right] = 2 \pi \hbar ( n + \frac{3}{4}) .
\end{equation}
Introducing $\nu$ and $z$, and identifying $L \equiv L_{sc}$, the equation takes the form,
\begin{equation}
\nu \left[ \sqrt{z^2-1} - \arccos \left( \frac{1}{z} \right) \right]
- \frac{\pi}{4} = \pi \left( n + \frac{1}{2} \right) ,
\end{equation}
that is just~(\ref{sem}).

\section{Calculation of the second order contribution to the quantization}
\label{appb}

The objective of this Appendix is to calculate the contour integral (\ref{oint2}). The contour must circle the turning point at $r_{min}$ and
also the point $r=R$ that represents the hard wall. It is convenient 
to compute this integral by taking the contour infinitesimally close
to the real line. The parts of the contour that are parallel to
the real r axis give the same real contribution, which can be computed
directly. Calculating this integral will reveal that the classical turning point gives an infinite contribution. The part of the contour that
circles the turning point also give infinite contribution, and the sum
of all the terms is finite.

 The standard way to deal with integrals of this kind is to integrate
and differentiate using a suitable parameter as a variable (usually the energy). Here one first integrates with respect to the energy. The number of integrations is such that the contribution from the turning point to the contour integral in the complex $r$ plane converges.
Then the contour integral is computed. The last step
is to differentiate with respect to the energy
to obtain the desired result. For this purpose~(\ref{sigma2}) is written in the form
\begin{equation}
\sigma'_2 = \frac{3 L_{sc}^2}{4 m r^3} \frac{\partial}{\partial E}
\frac{1}{(2 m E r^2 -L_{sc}^2)^{1/2}} - \frac{5 L_{sc}^4}{24 m^2 r^5}
\frac{\partial^2}{ \partial E^2} \frac{1}{(2 m E r^2 -L_{sc}^2)^{1/2}}
 - \frac{1}{8 r \sqrt{2 m E r^2  - L_{sc}^2} }
\end{equation}

Now each term will be treated separately. Integration of the first one
results in
\begin{equation}
I_1 = \int^R_{\sqrt{\frac{L_{sc}^2}{2 m E}}} \frac{3 L_{sc}^2}{2 m r^3}
\frac{dr}{\sqrt{2m E r^2 - L_{sc}^2}} = 
\frac{3}{4 m R^2} \sqrt{2m E R^2 - L_{sc}^2} - \frac{3 E}{2 L_{sc}}
\arcsin \left( \sqrt{ \frac{L^2_{sc}}{2 m E R^2}} \right) + \frac{3 E}{2 L_{sc}} \frac{\pi}{2} . 
\end{equation}

Differentiation with respect to E and change of variable from E to z gives:
\begin{equation}
\label{i1}
\frac{\partial I_1}{\partial E}= \frac{3}{2 L_{sc}} \frac{1}{\sqrt{ 
\frac{2 m E R^2}{L_{sc}^2}-1}} + \frac{3}{2 L_{sc}} \arccos
\left( \sqrt{ \frac{L^2_{sc}}{2 m E R^2}} \right) =  \frac{3}{2 L_{sc}}
\left( \frac{1}{\sqrt{z^2-1}} + \arccos \left( \frac{1}{z} \right) \right) .
\end{equation}

The next term can be treated in a similar manner,
\begin{equation}
I_2 = \int^R_{\sqrt{\frac{L_{sc}^2}{2 m E}}} \frac{5 L_{sc}^4}{12 m^2 r^5}
\frac{dr}{\sqrt{2m E r^2 - L_{sc}^2}} . 
\end{equation}
Substitution of $y=\frac{1}{r}$ helps to perform the integral and to obtain
\begin{equation}
I_2 = \frac{5 L_{sc}^2}{12 m^2} \sqrt{2 mE R^2 - L_{sc}^2} 
\left( \frac{3 m E}{4 L_{sc}^2 R^2} + \frac{1}{4R^4} \right) + 
\frac{5 E^2}{8 L_{sc}} \left( \frac{\pi}{2} - \arctan \left[
\frac{1}{\sqrt{ \frac{2m E R^2}{L_{sc}^2} -1}} \right] \right) . 
\end{equation}
This term has to be differentiated twice with respect to E, and then the
variable E will is replaced by z. After some manipulations the 
contribution of this term is found to be
\begin{equation}
\label{i2}
\frac{\partial^2 I_2}{\partial E^2} = 
\frac{5}{4 L_{sc}} \frac{1}{\sqrt{z^2 -1 }} - \frac{5}{12 L_{sc}}
\frac{1}{ (z^2 -1)^{3/2}} + \frac{5}{4 L_{sc}} \arccos \left(
\frac{1}{z} \right) .
\end{equation}

The third term is just
\begin{equation}
\label{i3}
I_3= \int^R_{\sqrt{\frac{L_{sc}^2}{2 m E}}} \frac{dr}{4 r \sqrt{2 m E r^2 - L_{sc}^2}} = \frac{1}{4 L_{sc}} \arccos \left( \frac{1}{z} \right) . 
\end{equation}

And the final result is
\begin{equation}
\oint \sigma'_2 dr = \frac{\partial I_1}{\partial E} -\
\frac{\partial^2 I_2}{\partial E^2} - I_3 = 
\frac{1}{4 L_{sc}} \frac{1}{\sqrt{z^2 -1}} + \frac{5}{12 L_{sc}} \frac{1}{ ( z^2 -1)
^{3/2}}
\end{equation}
that reduces to~(\ref{oint2}).

\section{Exact generalized angular momentum eigenvalues in WKB}

\label{appc}
The WKB series for the angular momentum in $D=3$ dimensions were constructed 
by
Salasnich and Sattin~\cite{salasnich97}. Summation of this series
leads to the correct form of the quantum angular eigenvalues, namely, $(l+1)l$.
Here this result will be generalized to the angular momentum in 
arbitrary dimension $D$.   

The first step is to compute the form of the Laplacian in Hyperspherical
coordinates. Then by separation of variables an equation relating the
generalized angular momentum operator to its projection on a $D-1$ dimensional
space is found. We will assume that the Laplacian has the form
\begin{equation}
 \Delta  = \frac{\partial^2}{\partial^2 r} + \frac{D-1}{r} \frac{\partial}{\partial r } + \frac{1}{r^2} \Delta^s_D 
\end{equation} 
where $\Delta^s_D $ is the angular part that does not depend on the radial
variable. 
We will use induction in the number of degrees of freedom to prove that the 
laplacian indeed has this form. First one should note that for $D=2$ and $3$ this
assumption holds. Assume it holds for a $D-1$ dimensional system and add an 
additional Cartesian coordinate $x_D$. 
The Laplacian is now
\begin{equation}
 \Delta = \frac{\partial^2}{\partial^2 x_D} + \Delta_{D-1}
\end{equation}
$\Delta_{D-1}$ is the Laplacian in the space of the coordinates $x_1,x_2,\cdots ,x_{D-1}$. Transforming  $x_1,x_2,\cdots ,x_{D-1}$ to Hyperspherical coordinates
and using the induction assumption we obtain
\begin{equation}
 \Delta  =  \frac{\partial^2}{\partial^2 x_D} + \frac{\partial^2}{\partial^2 r} + \frac{D-2}{r} \frac{\partial}{\partial r } + \frac{1}{r^2} \Delta^s_{D-1} 
\end{equation}

The transformation to Hyperspherical coordinates involves only the two
coordinates $r$ and $x_D$. All of the angular coordinates will not change and
therefore neither will $ \Delta^s_{D-1}$. One substitutes
\begin{eqnarray}
 x_D & = & R \cos \theta \nonumber \\
 r & = & R \sin \theta
\end{eqnarray}
The computation of the laplacian is straight forward and leads to
\begin{equation}
 \Delta = \frac{\partial^2}{\partial^2 R} + \frac{D-1}{R} \frac{\partial}{\partial R} + 
\frac{1}{R^2} \Delta^s_D
\end{equation}
where 
\begin{equation}
\label{angularoper}
 \Delta^s_D = \frac{\partial^2}{\partial^2 \theta} + (D-2) \cot \; \theta \frac{\partial}{\partial \theta} + \frac{1}{\sin^2 \theta} \Delta^s_{D-1}
\end{equation}
is the $D$ dimensional angular momentum operator.
Since the systematic high order WKB expansion works only for ordinary differential equations
we will assume that the exact form of the eigenvalues of the angular
momentum in lower
dimensions is known. The equation connecting eigenvalues of angular
momentum of $D$ and $D-1$ dimensions is obtained by separation of variables
\begin{equation}
 T'' (\theta) + (D-2) \cot (\theta) T'(\theta) - \frac{(m+D-3)m}{\sin^2 (\theta)}  T (\theta) = - \lambda^2 T (\theta) .
\end{equation}
where $T(\theta)$ is the eigenfunction for the eigenvalue $\lambda^2$, while
$m(m+D-3)$ is the eigenvalue of $\Delta^s_{D-1}$. It is justified by induction
in what follows (see~(\ref{exactangmom})).
This is the generalization of equation (4) in~\cite{salasnich97}. 
The computation follows~\cite{salasnich97} with only minor changes.
Instead of (5) in~\cite{salasnich97} one substitutes 
\begin{equation}
 T( \theta) = \frac{F (\theta)}{(\sin \theta )^{\alpha}} , 
\end{equation}
where $\alpha = \frac{D-2}{2}$. After some  manipulations and
substitution of $x= \theta + \frac{\pi}{2}$ one obtains
\begin{equation}
\label{known}
 - F'' (x) + \frac{U}{\cos^2 (x)} F (x) = E F (x) , 
\end{equation}
that is (7) of~\cite{salasnich97}. However, the values 
of the parameters are different. Here $ E = \lambda^2 + \alpha^2$
and $ U = m ( m + D -3) - \alpha ( 1 - \alpha)$. The WKB expansion   
of~(\ref{known}) is developed in~\cite{salasnich97,robnik97} for any order.
The quantization condition is given by a series which is then summed.
This results in
\begin{equation}
\label{summedseries}
\sqrt{E} - \frac{1}{2} \sqrt{1 + 4 U^2} = n + \frac{1}{2} , 
\end{equation}
But here $\sqrt{1 + 4 U^2} = 2 ( m + \frac{D-3}{2})$. Substitution of 
$U$ and $E$ leads to
\begin{equation}
\label{exactangmom}
 \lambda^2 = (n+m)( n + m + D-2)
\end{equation}
This is indeed the correct result for the generalized angular momentum
eigenvalues. The semiclassical eigenvalue is found from~(\ref{summedseries})
assuming $E \simeq \lambda^2$, leading to $\lambda \simeq n + m + \frac{D-2}{2}$.
Identifying $l=n+m$ leads to~(\ref{semiangmom}).
\end{appendix}

\typeout{References}


\begin{thebibliography}{99}

\bibitem{mount72}
{Berry M.V. and Mount K.E.},
{\em Rep. Prog. Phys.},
{\bf 35}, 315 (1972).

\bibitem{heading62}
{Heading J.}
{\em An Introduction to Phase-Integral Methods},
Methuen, London, (1962).

\bibitem{froman65}

{Fr\"{o}man N. and Fr\"{o}man P. O.},
{\em JWKB Approximation},
North-Holland, Amsterdam, (1965).


\bibitem{Dunham32}
{Dunham J. L.},
{\em Phys. Rev.},
{\bf 41}, 713 (1932).

\bibitem{bender77}
{Bender C.M., Olaussen K. and Wang P.S.},
{\em Phys. Rev. D},
{\bf 16}, 1740 (1977).

\bibitem{bender} 
{Bender C.M. and Orszag S.A.},
{\em Advanced Mathematical Methods for Scientists and Engineers},
McGraw-hill, New York, (1978).

\bibitem{gutz89}
{Gutzwiller M. C.}, {The semi-classical quantization of chaotic Hamiltonian
systems}, {In Giannoni M. J., Voros A. and Zinn-Justin J. ,editors},
{\em Proceedings of the 1989 Les Houches Summer School on 'Chaos and Quantum Physics'},
pages 201-249. Elsevier Science Publishers, Amsterdam, (1991).


\bibitem{primack97?}
{Primack H. and Smilansky U.}, 
{\em J. Phys. A:Math. Gen.},
{\bf 31}, 6253 (1998).

\bibitem{dalqvist99}
{Dahlqvist P.},
{\em chao-dyn 9812017},
{preprint}


\bibitem{prosen93} 
{Prosen T. and Robnik M.},
{\em J. Phys. A: Math. Gen.},
{\bf 26}, L37 (1993).

\bibitem{boasman94} 
{Boasman P.A.},
{\em Nonlinearity},
{\bf 7}, 485 (1994).

\bibitem{agam95} 
{Agam O.},
{\em Non-universal properties of chaotic systems},
{Ph. D. thesis},
Technion (1995).



\bibitem{robnik97}
{Robnik M. and Salasnich L.},
{\em J. Phys. A.:Math. Gen.}
{\bf 30}, 1711 (1997).
 
\bibitem{avery89} 
{Avery J.},
{\em Hyperspherical Harmonics},
Kluwer Academic Publishers, Dordrecht, (1989).

\bibitem{watson58} 
{Watson G.N.},
{\em A Treatise on the Theory of Bessel Functions.},
Cambridge University Press, (1958).



\bibitem{langer37}
{Langer R.E.},
{\em Phys. Rev.},
{\bf 51}, 669 (1937).

\bibitem{robnik97b}
{Robnik M. and Salasnich L.},
{\em J. Phys. A.:Math. Gen.}
{\bf 30}, 1719 (1997).

\bibitem{salasnich97}
{Salasnich L. and Sattin F.},
{\em J. Phys. A. :Math. Gen. },
{\bf 30}, 7597 (1997).

\bibitem{lazutkin93}
{Lazutkin V. F.}
{\em KAM Theory ans Semiclassical Approximations to Eigenfunctions},
Springer-Verlag, Berlin, (1993).

\bibitem{alonso93}
{Alonso D. and Gaspard P.}
{\em Chaos},
{\bf 3}, 601 (1993).

\bibitem{goldstein} 
{Goldstein H.},
{\em Classical Mechanics},
Addison-Wesley,
Reading MA, (1980).







\end{thebibliography}
\end {document}